\documentstyle[aps,epsfig]{revtex}
\begin{document}
\draft
\title{Piezoelectric mechanism of orientation of stripe structures in
two-dimensional electron systems}
\author{D.V.Fil}
\address{Institute for Single Crystals National Academy of Sciences of
Ukraine,
Lenin av. 60 Kharkov 61001 Ukraine\\
e-mail: fil@isc.kharkov.com}
\maketitle
\begin{abstract}
A piezoelectric mechanism of orientation of stripes in two-dimensional
quantum Hall systems in $GaAs$ heterostructures is considered.
The anisotropy of the elastic moduli and the boundary of the sample
are taken into account. It is found that in the average the stripes
line  up with the [110] axis.
In double layer systems the wave vector of the stripe structure rotates
from the [110] to [100] axis if the period of density modulation becomes
large than the interlayer distance. From the experimental point of view
it means that in double layer systems  anisotropic part of resistivity
changes its sign under variation of the external magnetic field.
\end{abstract}
\section{Introduction}
A homogeneous state of two-dimensional electron gas is unstable
at low densities and low temperatures. Under such conditions a transition
into a Wigner crystal state occurs. The minimum of the energy corresponds
to the triangular lattice \cite{1} for the case of classical Wigner crystal.
Recently, non-uniform electron states in quantum Hall systems have become a
subject of great attention. One can expect that many new states with a
spatial modulation of the electron density are realized in these systems.
In quantum Hall ferromagnets(QHF) skyrmions may form  lattice structures
\cite{2} (skyrmions in QHF obey the electric charge).
Skyrmions are spatially extended objects and the square
lattice is preferable at moderate densities. Among recent experimental
results the most intriguing one is the observation of strong anisotropy
of resistivity at filling factors $\nu =N+1/2$ ($N$ is the integer number,
$N\ge 4$)\cite{3,4}.
The effect is a signature of a stripe phase state of electrons of
the half-filled Landay level\cite{5,6}.

It has been found in experiments \cite{3,4} that the
low-resistivity direction is alinged with the [110] axis.
Therefore there is
an anisotropic electron-electron interaction
causes preferable orientation of the stripes
relative to the crystallography axes of the host material.
Such an interaction should be comparatively weak,
because the effect of rotation of the stripes in tilted magnetic fields
has been observed \cite{7,8}.

The modulated  electron structures emerge as a result
of competition between the direct Coulomb and exchange interaction.
In systems with cubic symmetry these interactions are isotropic.
They cannot cause the stripe to line up with a particular axis.
$GaAs$ is the piezoelectric crystal. Therefore one can suppose that
piezoelectric interaction (which remains anisotropic in cubic crystals)
plays the main role in orientation of stripe structures.

The anisotropy of the electron-electron interaction
in piezoelectrics was studied previously by Rashba and Sherman in
\cite{9}. They investigate the influence of piezoelectric interaction on
symmetry of Wigner crystal states. The isotropic model for
the elastic media was used in Ref. \cite{9}. The results \cite{9} cannot
be applied directly to $GaAs$ where the anisotropy of the elastic moduli is
rather strong.

In this paper we consider the piezoelectric mechanism
of orientation of the stripe structures in $GaAs$. We take into account the
anisotropy of the elastic moduli and the influence of the surface of
the sample. The major conclusion is that if the electron layer is
parallel to the (001) plane the energy of the stripe phase is minimal when
the angle $\phi$ between the wave vector of the stripe structure and the
[100] axis is in the interval $30^o\div 60^o$.
The potential curve as the function of $\phi$ shows a rather shallow shape
in this interval. The average direction
of the wave vector corresponds to the orientation
observed experimentally.

Double layer stripe structures are also considered in the paper.
We find that the stripes are aligned with the [100] or [010] axis,
if the period of the density modulation
becomes greater than the interlayer distance. The effect can be easily
checked experimentally, because the period of the stripe structure,
which is proportional to the magnetic length, is increased under lowering
of the external magnetic field.

\section{Monolayer system}

At the beginning we consider the electron layer in an infinite
piezoelectric cubic crystal.
The properties of the host system are described with three elastic moduli
$c_{11}$, $c_{12}$, $c_{44}$,
the dielectric constant
$\epsilon $ and one piezoelectric modulus $e_{14}$.
Here and below  only the case of a layer laying in the (001) plane is
analysed. The stripe structure is described as a charge density wave(CDW)
with the wave vector ${\bf b}$ directed at the angle $\phi$ to
the [100] axis:
\begin{equation}
\rho({\bf r})= \rho_0 \sin ({\bf b} {\bf r}_{pl}) \delta(z)
\label{1}
\end{equation}
where ${\bf r}_{pl}$ is the projection of ${\bf r}$ on the (001) plane.
The $z$ axis is along the [001] direction. The density of the energy of the
system is
\begin{equation}
F= {{\bf E D}\over 8 \pi } +{\sigma_{ik} u_{ik}\over 2}
\label{2}
\end{equation}
where
\begin{equation}
D_i = \epsilon E_i - 4\pi \beta_{i k l} u_{k l}
\label{3}
\end{equation}
is the dielectric displacement,
\begin{equation}
\sigma_{ik} =\lambda _{iklm} u_{lm} +\beta_{lik}E_l
\label{4}
\end{equation}
the stress tensor,
$u_{ik}$, the strain tensor,
${\bf E}$, the electric field,
$\lambda_{iklm}$ the elastic moduli tensor,
$\beta_{i k l}$, the piezoelectric moduli tensor
(in cubic crystals
$\beta_{i k l} =e_{14}/2$ at
 $i\ne k \ne l$ and $\beta_{i k l} =0$ otherwise).
The quantities ${\bf D}$ and $\sigma_{ik}$
satisfy the equations
\begin{eqnarray}
{\rm div} {\bf D} = 4 \pi \rho \cr
{\partial \sigma_{ik}\over \partial x_k} =0
\label{5}
\end{eqnarray}
Taking into account Eqs.(\ref{5})
and boundary condition for
$\sigma_{ik}$ ($\sigma_{ik} n_k =0$ at the free surface, ${\bf n}$
is the vector perpendicular to the surface)
we obtain for the total energy
\begin{equation}
E=\int d^3 r F = {1\over 2} \int d^3 r \rho({\bf r}) \varphi ({\bf r})
\label{6}
\end{equation}
where  $\varphi $ is the electric potential (${\bf E}=- \nabla \varphi$).
The value of $\varphi $ can be found from Eqs. (\ref{5}).
Rewriting Eqs.(\ref{5}) in terms of the Fourier-components of $\varphi $ and
the displacement ${\bf u}$ we obtain
\begin{equation}
M_{ik} V_k=Q_i,
\label{7}
\end{equation}
where
\begin{equation}
\hat{M}=\left( \matrix{\hat{\Lambda }&\hat{T}\cr
               -\hat{T}^+&{\epsilon q^2\over 4 \pi }}\right)
\label{8}
\end{equation}
with $\Lambda_{i k}=\lambda_{i k l m} q_l q_m$,
$T_i= -\beta_{ i k l } q_k q_l$,
\begin{equation}
V_i= \cases{u_{i {\bf q}}&i=1,2,3\cr \varphi_{{\bf q}}&i=4},\ \
  Q_i= \cases{0&i=1,2,3\cr \rho_{{\bf q}}&i=4}
\label{9}
\end{equation}
($\rho_{\bf q}$, the Fourier component of the electron density).
The solution of Eq.(\ref{7}) is
\begin{equation}
\varphi_{\bf q}= M^{-1}_{44}({\bf q}_{pl},q_z) \rho_{\bf q}
\label{10}
\end{equation}
(${\bf q}_{pl}$ is the projection of ${\bf q}$
on the (001) plane)).
Inserting $\varphi$ into Eq.(\ref{6}) yields
\begin{equation}
E = {\rho_0^2 S\over 8\pi } \int_{-\infty}^{\infty}
 d q_z M^{-1}_{44}({\bf b},q_z)
\label{11}
\end{equation}
($S$ is the area of the layer).
Taking into account the smallness of the piezoelectric constant,
the energy can  be written as the sum
\begin{equation}
E=E_c+E_{pe}^0+E_{pe}^{an}
\label{12}
\end{equation}
where
\begin{equation}
E_c={\pi \rho_0^2 S\over 2 \epsilon b}
\label{13}
\end{equation}
is the Coulomb energy,
$E_{pe}^0$, $E_{pe}^{an}$, the isotropic and anisotropic parts of the
energy of piezoelectric interaction between electrons.
The anisotropic part has the form
\begin{equation}
E^{an}_{pe}=  \chi E_c F(\phi)
\label{14}
\end{equation}
where $\chi=e_{14}^2/\epsilon c_{11}$  is the small parameter of the
expansion.
The function $F$  has the amplitude of order of unity. It depends on
the ratio between the elastic moduli.

If $c_{12}=c_{11}- 2 c_{44}$ (the case of the isotropic  elastic medium)
\begin{equation}
F(\phi)= A \cos 4\phi
\label{15}
\end{equation}
where
\begin{equation}
A= {9\pi \over 32}(1-{c_{11}\over 3 c_{44}})
\label{16}
\end{equation}
The answer Eq. (\ref{16}) was obtained in \cite{9} with another method.
Using the values of $c_{11}$, $c_{44}$ for $GaAs$ in
Eq. (\ref{16}) we obtain $A\approx 0.3$. The minimum of the energy reaches
at $\phi=\pi /4$. But if one can use the average values of the square
velocities of the longitudinal and the transverse sound instead of $c_{11}$,
$c_{44}$ correspondingly, the value of $A$ approaches to zero. It means that
even the sign of the anisotropy remains unsettled. Thus it is a principal
question to take into account the anisotropy of the elastic moduli.

\begin{center}
\begin{figure}
\centerline{\epsfig{figure=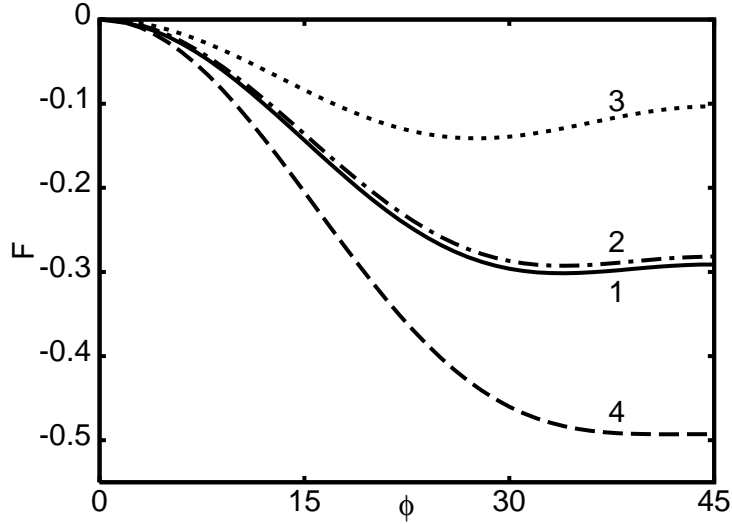,width=12cm}}
\vspace{0.5cm}
\caption{
The dependence of the energy
(in units of $\chi E_c$)
on $\phi$ (in degrees) ($\phi$ is the angle between the wave
vector of the stripe structure and the [100] direction).
Curve 1  corresponds to the infinite system, curves 2,3,4 -
the systems with the boundary.
2- $d/a=0$; 3 - $d/a=0.15$;
4 - $d/a=0.5$ ($d$ is the distance between the layer and the surface
of the sample, $a$, the period of the stripe structure).}
\label{fig1}
\end{figure}
\end{center}

The integral in Eq.(\ref{11}) has been computed numerically for the
anisotropic case. The values of the elastic moduli for
$GaAs$ ($c_{11}=12.3$, $c_{12}=5.7$, $c_{44}=6.0$, all in
$10^{11}$ dyne/cm$^2$) are used.
The result for $F(\phi)$ is shown in Fig. \ref{fig1} (curve 1).
We see from the
dependence obtained that the minimum of the potential energy
corresponds to  $\phi \approx 30^o$.  But the energy variation in the
interval $30^o<\phi <60^o$ is very small and all the configuration
in this interval are realized with an almost equal probability.
In the average the wave vector of the stripe structure is directed
along the  [110] axis. The absolute value of the anisotropic part of the
energy is
determined by the parameter $\chi$. For $GaAs$
($e_{14}=0.15$ C/m$^2$, $\epsilon=12.5$)
$\chi\approx 2\cdot 10^{-4}$.

An important question is how the result is changed under small variation
of the elastic moduli. We compute the function $F(\phi)$ versus the
$c_{12}$ modulus. If the value of $c_{12}$ is decreased the local maximum
at $\phi=\pi /4$ transforms to the global minimum (at $c_{12}\approx
5\cdot 10^{11}$ dyne/cm$^2$). In the opposite case
(increasing of this modulus) the minimum at $\phi\approx 30^o$ becomes
dipper. The situation in $GaAs$ is just in between these two cases.

The dependence of the energy of interaction on the angle $\phi$ remains
unchanged for the stripe structure with density modulation of
general form.
If one replaces Eq.(\ref{1}) with the sum over
the wave vectors ${\bf q}_n=n {\bf b}$ ($n$ is integer),
the quantities $E_{c}$ and $E_{pe}^{an}$
are multiplied with the same factor and the function $F(\phi)$ is
not modified.

In case of the electron crystal with square lattice
the energy can be written as a sum over the reciprocal lattice vectors
and each term in the sum depends
on the direction of the corresponding  wave vector.
Therefore a concrete form of the density distribution should be specified
to compute the energy. If the electron crystal is modelled as two CDW with
perpendicular  wave vectors the preferable orientation and the anisotropic
energy are obviously the same as for the stripe structure.
The anisotropic energy is much smaller for the triangular lattice
(modelled as the sum of three CDW).
The absolute value of the anisotropic energy is in $\approx 10^2$
times smaller than in the stripe state.
The minimum reaches when the angle between the [100] axis and one of
the CDW wave vectors is $k\pi /6$ ($k$ is the integer number).
Note that anisotropic contribution into the total energy
of the triangular lattice does not emerge in the isotropic elastic
medium (see Eq.(\ref{15})).

A finite thickness of the electron layer can be taken into consideration
in the approach used. The corresponding form factor
should be added into Eq. (\ref{11}). However, since the thickness of the
layer is much smaller than the period of the stripe structure,
it does not result in essential changes.

Usually the electron layers in heterostructures are situated close
to the surface of the sample ( a typical distance $d$ between the layer
and the surface is $\approx 5\cdot 10^3 \AA$). Therefore a principal
question is how  the influence of the surface may change the result
obtained.
To answer the question the system (\ref{5}) should be
solved  with taking into account the boundary conditions.

The solution of (\ref{5}) can be written in the form
\begin{eqnarray}
u_i= u_i(z) {\rm  e}^{{\rm i} {\bf b r}_{pl}}+ c.c. \cr
\varphi= \varphi(z) {\rm  e}^{{\rm i} {\bf b r}_{pl}} + c.c.
\label{17}
\end{eqnarray}
where $u_i(z)$ and $\varphi(z)$
satisfy the system of differential equations
\begin{eqnarray}
(c_{44}(\partial_z^2-b_y^2)-c_{11}b_x^2) u_x - (c_{12}+c_{44}) b_x (b_y u_y -
{\rm i}  \partial_z u_z) - {\rm i} e_{14} b_y \partial_z \varphi =0 \cr
(c_{44}(\partial_z^2-b_x^2)-c_{11}b_y^2) u_y - (c_{12}+c_{44}) b_y (b_x u_x -
{\rm i}  \partial_z u_z) - {\rm i} e_{14} b_x \partial_z \varphi =0 \cr
(c_{11}\partial_z^2 - c_{44}b^2) u_z +{\rm i} (c_{12}+c_{44})
\partial_z (b_x u_x + b_y  u_y) +  e_{14} b_x b_y \varphi =0 \cr
\epsilon(\partial_z^2-b^2)\varphi +4\pi e_{14}
({\rm i} \partial_z (b_x u_y +b_y u_x) -b_x b_y u_z)=0
\label{18}
\end{eqnarray}
with the boundary conditions
\begin{eqnarray}
\sigma_{iz}|_{z=d-0}=0\cr
\varphi |_{z=d-0}=\varphi |_{z=d+0}\cr
D_z |_{z=d-0}=(\partial_z\varphi)|_{z=d+0}\cr
\sigma_{iz}|_{z=-0}=\sigma_{iz}|_{z=+0}\cr
\varphi  |_{z=-0}=\varphi  |_{z=+0}\cr
u_i |_{z=-0}=u_i |_{z=+0}\cr
D_z |_{z=+0}- D_z |_{z=-0}= -2 \pi {\rm i}\rho_0 \cr
\label{19}
\end{eqnarray}
In Eq. (\ref{19})
\begin{eqnarray}
\sigma_{x(y) z}=c_{44}(\partial_z u_{x(y)} + {\rm i} b_{x(y)} u_z)-
{\rm i}{e_{14}\over 2} b_{y(x)} \varphi \cr
\sigma_{z z} = c_{11}\partial_z u_z + {\rm i} c_{44} (b_x u_x + b_y u_y)\cr
D_z = -\epsilon \partial _z \varphi - {\rm i} 2 \pi  e_{14}
(b_x u_y + b_y u_x)
\label{20}
\end{eqnarray}

An analytical solution of (\ref{18}) can be written if
the roots of the characteristic equation are found.
The roots were computed numerically.
The energy has been  separated similar to Eq.(\ref{12}),
with the Coulomb part
\begin{equation}
E^{'}_c={\pi \rho_0^2 S\over 2 \epsilon b}
(1+{\epsilon -1\over \epsilon +1} {\rm e}^{-2 b d})
\label{21}
\end{equation}
and the anisotropic part
\begin{equation}
E^{an}_{pe}=  \chi E^{'}_c F(\phi)
\label{22}
\end{equation}

The dependences $F(\phi)$ obtained are shown in Fig. \ref{fig1}
at several values
of the parameter
$d/a$ ($a=2\pi /b$ is the period of the
stripe structure).
At $d/a=0$ (the electron layer is situated on the surface of the sample)
the dependence (curve 2) is close to the one obtained for the
infinite medium (curve 1).
Under increasing of the ratio $d/a$ the minimum at
$\phi\approx 30^o$ becomes dipper (curve 3). But the shape of the potential
remains shallow in the interval $30^o<\phi <60^o$.
Then the reverse changes begins and
at $d/a\approx 0.5$
the two wall structure of the potential near $\phi=\pi /4$
disappears (curve 4). The anisotropy is maximal in the
last case.
The dependence approaches the curve 1
under further increasing of the parameter $d/a$
Thus, the surface of the sample does not influence essentially
on the orientation of the stripes.
The results obtained show that the piezoelectric mechanism
describes the experimental situation correctly.

\section{Double layer system}

In this section we consider the piezoelectric mechanism of orientation of
the stripe structures in double layer systems.
Double layer electron systems are widely used in experiments.
Therefore it is of interest to extend our consideration to the case
of two identical stripe structures in the adjacent layers.
There is another more important reason to address this problem.
In the double layer system there is an additional parameter -
the ratio between the interlayer distance and the period of the
stripe structure. The period is proportional to the magnetic length.
Therefore this parameter can be easily changed experimentally.
One can hope that the orientation of the stripes depends on this parameter.
If it is the case, the dependence can be extracted from the resistivity
data. Comparing the experimental data with the theoretical prediction
one can verify the model under consideration.

Basing on the results of the previous section we consider only the
infinite system. In the double layer system the Coulomb interaction results
in a relative shift of CDW in the adjacent layers. The electron density reads
as
\begin{equation}
\rho({\bf r})= \rho_0 \sin ({\bf b r}_{pl}) (\delta(z-{s\over 2})-
\delta(z+{s\over 2}))
\label{23}
\end{equation}
($s$ is the interlayer distance).
Following the steps of the previous sections we find
\begin{equation}
E = {\rho_0^2 S\over 4\pi } \int_{-\infty}^{\infty}
 d q_z M^{-1}_{44}({\bf b},q_z)      (1-\cos (q_z s))
\label{24}
\end{equation}

If the elastic moduli are isotropic
\begin{equation}
E_{pe}^{an}=  \chi E_c A \cos 4\phi
\label{25}
\end{equation}
where
\begin{equation}
A=2 \int_{-\infty}^{\infty}
 d z  {1-\cos (z s b)\over (1 + z^2)^4 }
({c_{11}\over c_{44}} - z^2 (8 {c_{11} \over c_{44}}  + 9))
\label{26}
\end{equation}
Calculating the integral in Eq.(\ref{26}) yields
\begin{eqnarray}
A= {9 \pi \over 16} (
1-{c_{11}\over 3 c_{44}}  -{\rm e}^{-s b}
((1+s b)(1-{c_{11}\over 3 c_{44}}) +
(s b)^2
({2 c_{11}\over 3 c_{44}}- {s b\over 3} (1-{c_{11}\over c_{44}}))))
\label{27}
\end{eqnarray}

\begin{center}
\begin{figure}
\centerline{\epsfig{figure=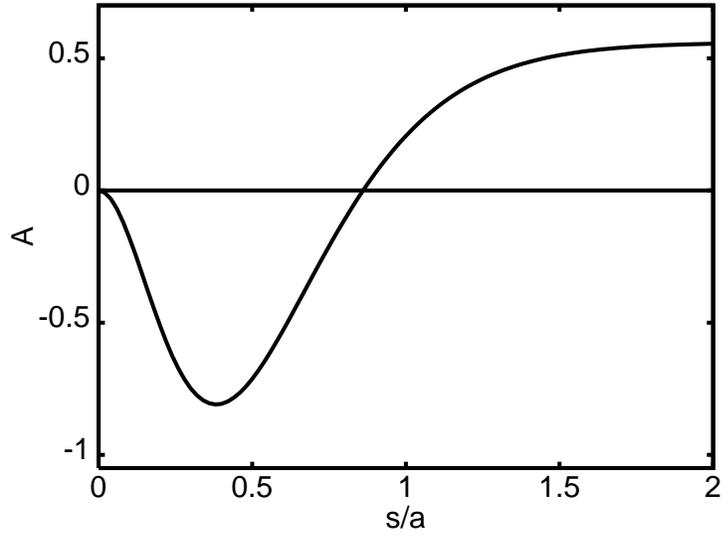,width=12cm}}
\vspace{0.5cm}
\caption{The dependence of the amplitude of the anisotropic energy
on the interlayer distance
for the double layer system in an isotropic elastic medium}
\label{fig2}
\end{figure}
\end{center}

The dependence of $A$ on the parameter $s/a$
at $c_{11}/c_{44}=12.3/6$ is shown in Fig. \ref{fig2}
One can see that at $s/a<1$ the anisotropic part of the energy
changes the sign and the stripes line up with the [100] or [010] direction.

\begin{center}
\begin{figure}
\centerline{\epsfig{figure=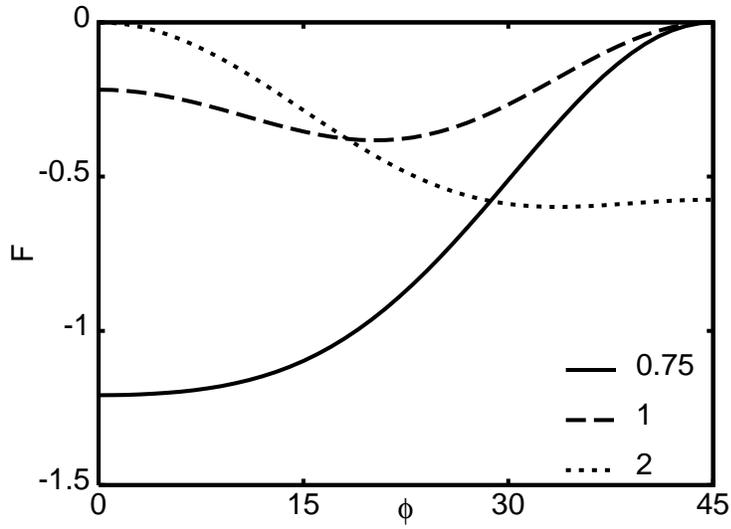,width=12cm}}
\vspace{0.5cm}
\caption{
The dependence of the anisotropic energy on the stripe direction
for the double layer system in $GaAs$
(solid curve, $s/a$=0.75; dashed curve,$s/a=1$;
dotted curve,$s/a=2$)}
\label{fig3}
\end{figure}
\end{center}

\begin{center}
\begin{figure}
\centerline{\epsfig{figure=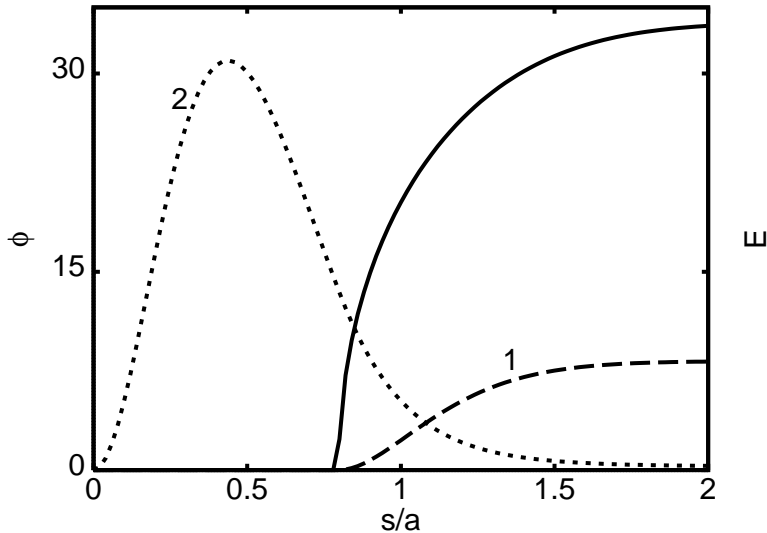,width=12cm}}
\vspace{0.5cm}
\caption{Orientation and the anisotropic energy versus the interlayer
distance in $GaAs$.
Solid curve, $\phi$ (in degrees);
dashed curve, the energy gain (in relative units) relative
the $\phi =0$ phase;
dotted curve, the energy gain  relative
the $\phi =\pi/4 $ phase.}
\label{fig4}
\end{figure}
\end{center}

Anisotropic systems show similar behavior.
For $GaAs$ the dependence of the energy on the angle $\phi $
is shown in Fig. \ref{fig3} at several values of $s/a$.
The angle $\phi$ which corresponds to the state with minimal energy
versus the parameter $s/a$ is shown in Fig. \ref{fig4}.
The energy gain
of the lowest energy state  comparing to
the $\phi=0$  (curve 1)
and  $\phi=\pi/4$ states (curve 2)
is also presented in Fig. \ref{fig4}.
One can see from Figs. \ref{fig3}, \ref{fig4}
the interaction between the layers does not
change the orientation of the stripes at $s/a>1.5$.
At  $0.8<s/a<1.5$ the stripe phases with $\phi\ne 0, \pi/4$
can be detected.
At $s/a<0.8$ the stripes line up with the [100] or [010]
direction.

The effect obtained is of great importance. The  parameter
$a$ depends on the magnetic length and the stripe phases at
$\nu =N+1/2$ have different periods at different $N$.
In the double layer systems
the stripes will have different orientation versus the
filling factor (along [110] at smaller $N$ and along [100]
at larger $N$). The prediction can be verified experimentally.
The effect if observed will be the experimental evidence of the
major role of the piezoelectric mechanism of orientation of
the stripe structures. It can also be used for indirect measurements
of the periods of the stripe structures.

\section{Conclusions}

We have shown that the piezoelectric interaction may play an essential
role in orientation of non-uniform two-dimensional electron
structures relative to the crystallography axes of the host material.

The anisotropic part of the energy of the stripe structure
in the quantum Hall system in $GaAs$ heterostructures
has been found.
If the electron layer is parallel to the (001) plane
the minimum of the energy corresponds
to the wave vector of the stripe structure directed at the angle
$\phi\approx 30^o$ to the [100] axis. The potential energy shows
a rather shallow shape at $30^o<\phi<60^o$ and all the direction in
this interval can be realized with almost the same probability.
In the average the stripes line up with the [110] axis.

The influence of the surface of the sample on the orientation of the stripes
has been analysed. If the electron layer is situated on the surface of
the sample
(the distance $d$ between the layer and the surface is much smaller than the
period $a$ of the stripe structure) the anisotropic part of the energy
is almost the same as for the infinite system. If the ratio $d/a$ is of
order of unity some changes may take place. At $d/a\approx 0.5$ the
local maximum at $\phi=\pi /4$ transforms into the global minimum.
At $d/a\approx 0.15$ the minimum at $\phi\approx 30^o$ becomes dipper and
the orientation of stripes along a low symmetry direction can be observed
at very low temperatures.

The results obtained describes the experimental situation correctly.
The absolute value of the anisotropic energy is small ( in $\approx 10^4$
times smaller than the Coulomb energy). Therefore external fields
(for instance, the tangent magnetic field) may change
the orientation (this effect observed experimentally).

The influence of the piezoelectric interaction on  the orientation of
the stripes is investigated with reference to the double layer systems.
For such systems the model predicts that at small interlayer distances
the stripes change their direction and line up with the [100] axis.
The re-orientation takes place if the interlayer distance becomes smaller
than the period of the stripe structure. The prediction
can be checked experimentally since the period is proportional to the
magnetic length, e.g. it will change under magnetic field variation.

If the effect is observed it can be
considered as the experimental evidence of the major role of the
piezoelectric mechanism of orientation of stripe structures.

This work was supported in part by the INTAS grant No. 97-0972.

\end{document}